\documentclass[conference]{IEEEtran}

\usepackage{cite}
\usepackage{amsmath,amssymb,amsfonts}
\usepackage{algorithmic}
\usepackage{graphicx}
\usepackage{textcomp}
\usepackage{xcolor}
\usepackage{comment}
\usepackage{url} 
\def\BibTeX{{\rm B\kern-.05em{\sc i\kern-.025em b}\kern-.08em
    T\kern-.1667em\lower.7ex\hbox{E}\kern-.125emX}}
    
\usepackage{booktabs}

\begin{document}

\title{The Uneasy Marriage of AI and Dependability \\ {\Large Integrating Taxonomy and Methods for Dependability and Accuracy Enhancement}}

%\author{Identity of authors hidden for double-blind review}

\author{
\IEEEauthorblockN{Aad van Moorsel}
\IEEEauthorblockA{\textit{School of Computer Science}}
\textit{University of Birmingham}\\
Birmingham, United Kingdom \\
a.vanmoorsel@bham.ac.uk %\\
%0000-0001-7233-6943
}

\maketitle

%\renewcommand{\shortauthors}{Undisclosed for review}

\begin{abstract}
In this paper we discuss the connection between fault-tolerance mechanisms in traditional computer systems, and approaches in accuracy enhancement for AI-based services.  We will find that AI mechanisms such as ensembles and reject option have direct counterparts in hardware and software dependability through N-modular redundancy and acceptance tests, even though their motivation, justification and implementation are quite different.  We augment the traditional dependability taxonomy to include critical defining features of faults and failures in AI-based services.  We propose to consider incorrect outcomes from AI as errors, even if the system hardware and software operates error free.   AI then becomes a third system layer (after hardware and software) for which dependability needs to be considered, and for which dependability has specific characteristics. The existing fault classes in the dependability taxonomy are not suited for AI, and we propose to introduce AI Output Faults, representing the inherent possibly incorrect (and therefore faulty) outcome of AI algorithms. We then map and compare fault tolerance mechanisms with AI accuracy enhancement mechanisms, and we see they carry strike resemblances. We hope the work presented in this paper will help in establishing a truly integrated and unified understanding of dependability for modern-day AI-based systems.    
\end{abstract}

% \caption{Seattle Mariners at Spring Training, 2010.}
% seats. Ichiro Suzuki preparing to bat.}


\section{Introduction}
\label{s:intro}
There has always been an uneasy marriage between the systematic dependability design approaches and the use of AI in systems that have high or stringent dependability requirements. This is unfortunate, since the systemic approach that underpins dependability design is an unmissable tool in providing confidence that the design considers all main failure types, utilises the most powerful fault tolerance mechanisms, and does not get itself lost in details that may be relevant for implementation but do not principally matter to design decisions. Therefore, this paper tries to bridge dependability design approaches with AI approaches that improve accuracy of AI-based services, in the hope that the result help develop a unifying perspective relevant to both dependability and AI system designers.  

We take a systematic, back to basics, approach to unifying AI and dependability thinking.  We start from the dependability framework and taxonomy developed in the seventies and eighties \cite{Avizienis04}, and integrate AI-based systems into this framework.  Then we consider and compare dependability enhancing mechanisms for traditional systems and accuracy enhancing mechanisms for AI-based systems.  To be precise, throughout what follows, the system under consideration is a service implemented through an AI algorithm (at run-time, that is, the training phase was completed), and our notion of failure equates to an incorrect output (for a given input). 

AI safety has gained a lot of attention in recent years, which is directly relevant to the dependability framing in this paper.  Reviewing all recent developments would distract from the purpose of this paper. Arguably much of that attention has gone to the challenges we may face in terms of societal impact, ethics, impact on work, and even the ruin of the world as we know it.  As pointed out when discussing the `functionality fallacy' \cite{Raji2022}, this may have been at the detriment of considering building functionally correct services, and this paper want to return to the topics of building dependable AI-based systems. In this context, a main contribution has been made by the many-faceted discussion of dependable AI service in \cite{bloomfield2024dependability}.  We also point to the recent paper on whole-system dependability by Vieira \cite{Vieira2025TrustSystemsNotJustAI}, but we are not aware of a systemic fitting of AI within the dependability taxonomy and associated fault tolerance mechanisms. This paper builds on an informal presentation and article about fault tolerance mechanisms and their AI counterparts \cite{moorsel25}, which discusses some of the techniques examined in Section \ref{s:FT}. 

The dependability taxonomy paper \cite{Avizienis04} does not mention AI-based services, although it anticipates to require adjustments, for instance for systems with emerging properties.  Since then, AI has penetrated all services, from generative AI interfaces, to business-critical applications in analysis, trading or security systems, and in safety-critical applications such as autonomous vehicles. Particularly the latter two categories make it necessary to consider the dependability of the system--we argue that it makes it necessary to introduce 'dependability thinking' into the design of AI-based services and systems.  To deal with AI in the dependability framework, we suggest in Section \ref{s:dependability} to consider AI components as separate from hardware or software, introducing specific types of faults not found in traditional hardware or software.  We then consider all the fault types in \cite{Avizienis04}, and explain that AI faults fit categories that the dependability framework previously dismissed as `poor decision making' and the like.  We therefore introduce basic fault types to cover AI-based systems, starting from AI Faults and AI Output Faults, as we discuss in Section \ref{s:dependability}.  

In addition, we compare traditional fault tolerance mechanisms with the approaches available to improve the accuracy of AI systems during run-time.  It turns out that these two sets of mechanisms contain a lot of similarities: the acceptance test in recovery blocks in fault tolerance are related to the reject option, combined with a form of cascades, in AI, and N-modular redundancy in fault tolerance has its counterpart in AI ensembles.  Errors in the AI output are unavoidable but also undetectable (otherwise one would not need the AI to begin with).  We will see that the oracle functionality, which in dependability manifests itself in acceptance tests and error signaling, in AI is usually implemented through using the values of model parameters, even if their semantic meaning for system dependability is not always clear.  Finally, we will touch on metrics for dependability and accuracy, respectively, which are different and whose relationship is not always obvious. Therefore, to determine the dependability of AI-based systems one would need to relate accuracy metrics (for instance expressed in probabilities of false and true positive and negatives) with dependability metrics such as reliability. 

The tension between dependability and AI, which in the title is called the `the uneasy marriage', may merit a brief philosophical discussion. Both dependability design and AI design start from the realisation that no system is perfect. In AI this is straightforwardly obvious, since AI-based services cannot be fully correct, irrespective of the amount of data and training one performs before deployment. AI thinking focuses on improving the accuracy, often expressed as SOTA or state of the art.  Dependability is based on a similar realisation, that it is not possible to design out and avoid all faults, and that fault tolerance mechanisms are required, even for the best built systems.  Dependability thinking focuses on reducing the impact of failures. These two perspectives mirror each other, one being accustomed to take a positive view, and celebrating any improvement in accuracy despite it remains far from perfect.  The other (dependability) more used to a pessimistic view, being concerned about how to deal with any possible error, despite the already highly reliable nature of the overall systems.  The extreme form of the positive perspective is that AI will resolve any dependability challenges faced by AI, the extreme form of the negative perspective is that AI should never be used for dependable systems since it inherently is faulty.  We believe that these opposite viewpoints have led to the two communities staying farther apart than desirable.  Instead, we believe that there is a lot to gains from designing based on integrated and unified considerations of AI and dependability, and this paper contributes to bridging between these disciplines. 

\section{AI and the Dependability Taxonomy}
\label{s:dependability}
In this section we aim to integrate AI within the dependability taxonomy and framework, and we discuss the challenges we face if so doing. We focus on errors in outputs of AI algorithms, since in this respect AI is different from faults traditionally considered.  

{\bf Background.\ } The seminal paper by Avizienis and colleagues, which opened the inaugural issue of IEEE Transactions on Dependable and Secure Systems \cite{Avizienis04}, provides a summary of dependability terminology and the systemic approach to designing and evaluating fault tolerant systems. It introduces the fault $\rightarrow$ error $\rightarrow$ failure trifecta: the {\em fault} is the underlying reason, usually latently present in the system, for a possible failure.  The fault can be a software bug, or the hardware that is susceptible to bit flips. Such a fault may lead to an {\em error}, which is the observable manifestation of something going wrong. However, this error might not yet equate to the {\em failure} of the system under consideration (e.g., the error could have been a component failure), and could still be handled through a recovery mechanism.  Failure occurs when the system stops functioning.  The notion of failure thus depends on which system one has considers, and one should carefully consider the system boundaries that delivers the service one wants to be dependable. 

The term dependability has one obvious definition \cite{Avizienis04}, namely that the system should not exhibit more than acceptable service failures.  This is rather obvious as a definition, but of particular interest is the traditional definition of dependability, also available in \cite{Avizienis04}, which says that a system is dependable if one can justifiably rely on the service it delivers. The latter may be considered more convoluted, but the element of justification is particularly relevant when considering AI.  One does not only want to avoid failures, one would actually want to have evidence of the dependability of the system, so that trust in the AI-based service is justified. 

\begin{figure}[htbp]
  \centering
  \includegraphics[width=0.45\textwidth]{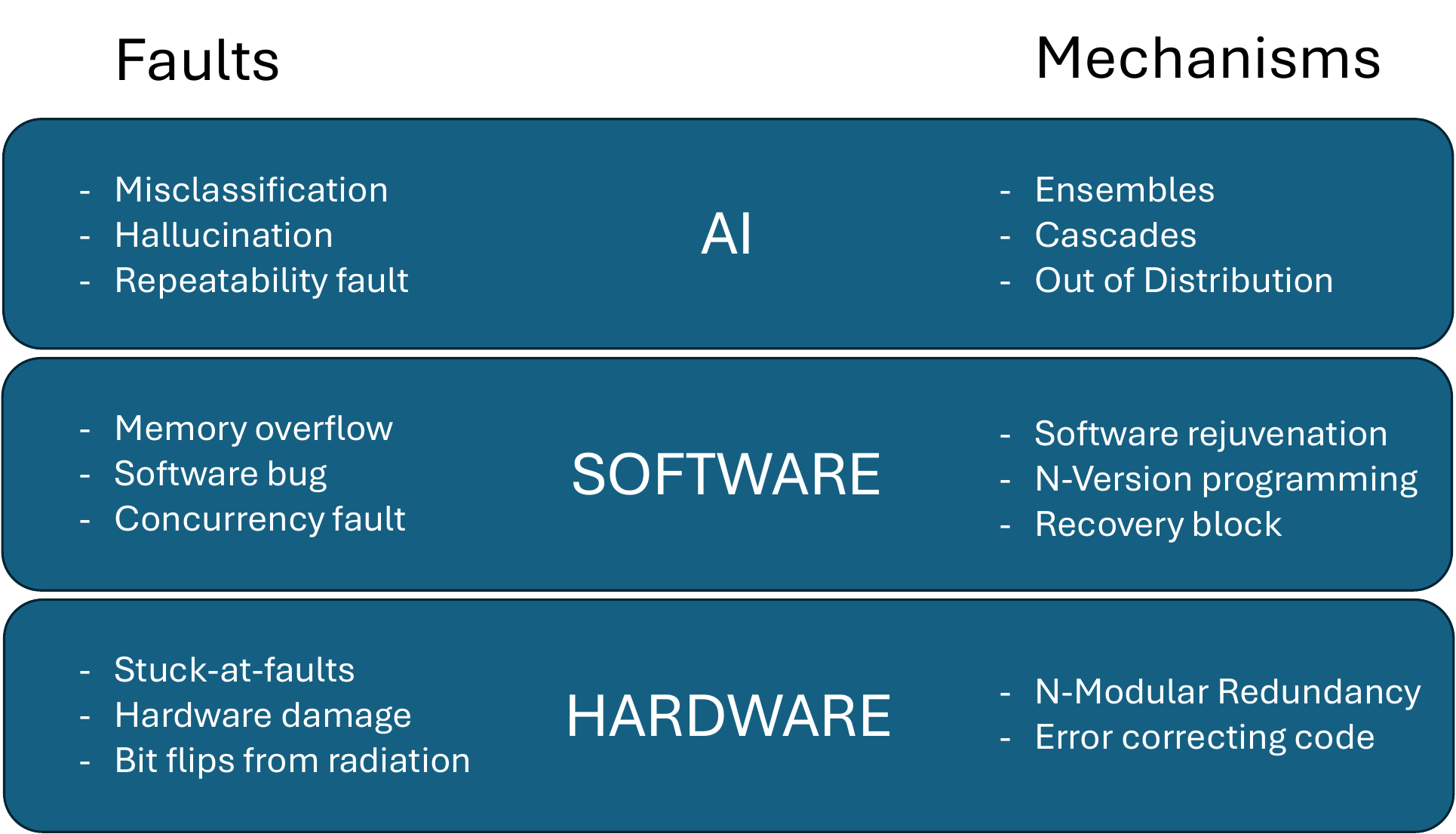}
  \caption{Separate consideration of AI faults and mechanisms from hardware and software dependability.}
  \label{fig:trifecta}
\end{figure}

{\bf Separate Treatment of AI-based Services.\ } We argue that the nature of faults in AI is substantially different from that in software and hardware, and that it is useful to consider them separately (as depicted in Fig. \ref{fig:trifecta}).  Consider an AI-based service that uses a binary classifier.  The simple example is a binary classifier that distinguishes between cats and dogs, or recognises stop signs in autonomous vehicles.  If the picture is that of a cat, then it is not unreasonable to consider it an error if it classifies it as a dog.  The fault that causes the error is simply the classifier, irrespective of whether it was implemented completely correctly and the hardware is completely error free.  In other words, the fault is a fault by design.  Although dependability is based on a related realisation, namely that a system without any faults is an illusion, the notion that we introduce the fault in software knowingly, and cannot remove it contributes to the uneasy marriage between AI and dependability. 

Acceptance of a notion like fault by design is difficult to square when designing safety-critical systems such as autonomous vehicles, but this is effectively the case in modern-day design if autonomous systems.  Some hardware faults, eg associated with radiation, are equally unavoidable (although not knowingly introduced), and may therefore be considered similar to AI related faults, but these are not knowingly introduced, and at the least these are extremely rare to lead to errors or failures.  In AI, inaccurate outcomes are part and parcel of the use of AI technology.  Note that no amount of training or data can avoid failures, and one should assume that AI errors will occur unavoidably \cite{bloomfield2024dependability}. 

\begin{figure}[htbp]
  \centering
  \includegraphics[width=0.45\textwidth]{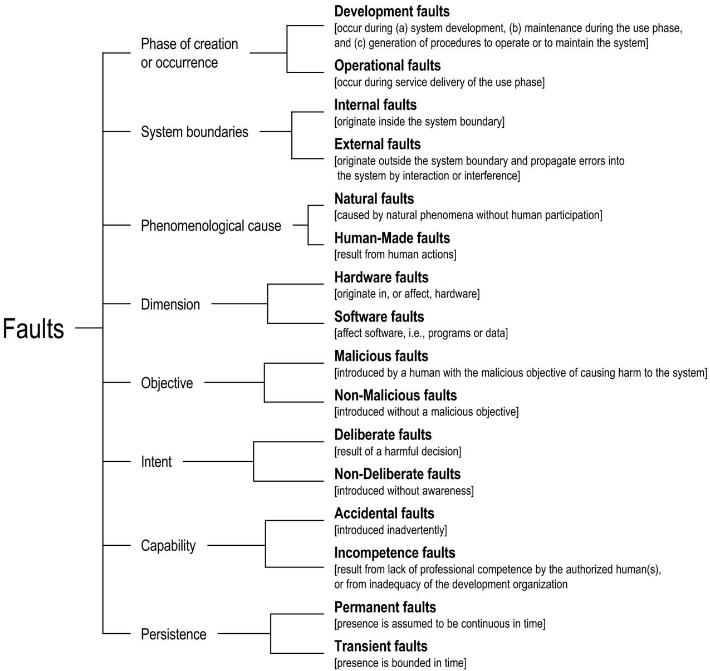}
  \caption{From \cite{Avizienis04}, fault classes.}
  \label{fig:faultclasses}
\end{figure}

{\bf AI Fault Types.\ } The fault classes identified in Section 3 of \cite{Avizienis04} and copied in Fig. \ref{fig:faultclasses} exemplify shortcomings when applying this taxonomy to AI-related faults.  There are 8 binary fault classes and \cite{Avizienis04} already predicts that these may not be exhaustive, but as we will see there are not only missing types, but also a few misconceptions in the classification. Traversing down the fault classes of Fig. \ref{fig:faultclasses} we obtain the following for faults in the outcomes of AI algorithms: 
\begin{itemize}
    \item {\em Phase of creation or occurrence}: {\em Development Fault}
    \item {\em System boundaries}: {\em Internal Fault}
    \item {\em Phenomenological cause}: {\em Human-Made Fault}
    \item {\em Dimension}: {\em Software Fault}
    \item {\em Objective}: N{\em on-Malicious Fault}
    \item {\em Intent}: {\em Deliberate Fault}
    \item {\em Capability}: {\em Incompetence Fault}
    \item {\em Persistence}: {\em Permanent Fault}
\end{itemize}

{\bf AI Faults.\ } With respect to {\em Dimension}, we argue that the type of fault stemming from AI algorithms is neither hardware or software.  The natural resolution to this shortcomings is to add under {\em Dimension} the subclass of {\em AI Faults} in addition to {\em Software Faults} and {\em Hardware Faults}.  One may argue that this destroys the binary nature of all faults classes (each class has two subclasses), although this may not be important in the bigger scheme.  

Regarding the fault class {\em Intent}, the incorrect outcome of an AI algorithm is a deliberate fault, but cannot be called 'result of a harmful decision'.  A better terms may be {\em Conscious Fault} instead of {\em Deliberate Fault}, and the term 'harmful' in the description can be replaced by 'conscious'.  Regarding {\em Capability}, AI faults certainly are not {\em Accidental Faults}, but to categorise them as {\em Incompetence Fault} is not appropriate.  One can replace the term {\em Incompetence}, so that this fault class covers AI as well. (The text in Fig. \ref{fig:faultclasses} is amusing and may find some supporters: `result from lack of professional competence by the authorized human(s), or from inadequacy of the development organisation'.)  

It should be noted that not every class of fault needs to have a meaningful choice for each of the nine fault classes in Fig. \ref{fig:faultclasses}, one could choose the {\em Capability} class to be not specifically relevant for AI-based systems.  If one wants to avoid adding a third category {\em AI Faults} under {\em Dimension}, one could instead decide that {\em Dimension} is not relevant to AI output related faults, or decide that AI faults fall under {\em Software Faults}.  Instead one can introduce the class {\em Output Fault} with the binary classification {\em Deterministic Fault} (for non-AI systems, simply a wrong output) and {\em Uncertainty Fault} (for AI systems, where the outcome is sometimes wrong because of the inherent uncertainty in AI-based systems).  However, we believe it is more natural and useful to introduce {\em AI Faults} under {\em Dimension} to stress that faults in AI outputs are of different nature than those caused by software or hardware.  

When considering the outcomes of AI algorithms, generative AI outcomes important to consider.  In generative AI, the notion of correct outcomes is yet harder to identify then for other algorithms such as classifiers.  For classifiers, we would be able to distinguish correct from incorrect in a manner similar to output from other computer systems.  This does not mean this assessment can be automated, but in principle we are able to reason about correct or not correct.  In generative AI, whether the output is correct or not is not defined, and up to interpretation and knowledge by an oracle (e.g., the user, but for the more straightforward cases, the oracle is increasingly a RAG, a Retrieval-Augmented Generation component).  For this reason, we would want to add a fault class {\em AI Output Fault}, with as options {\em Non-ambiguous Fault} and {\em Semantic Ambiguity Fault}.  Similarly, we can make a case for identifying the challenge to assure that outcomes are identical for the same input as a {\em Repeatability Fault} and the fact that outcomes are sometimes hallucinated, a {\em Hallucination Fault}. 

{\bf Other AI-related Faults.\ } The AI fault classes discuss thus far all relate to the situation in which the use of AI results in uncertainty or errors because of the inherent incorrect outputs all AI algorithms exhibit.  That is, these relate to the situation in which the algorithms executes as well as possible and is `functionally correct'.  How about if there are other faults that contribute to increasing the error of the output?  

During the learning phase, there could be all types of faults present within the learning process.  For instance, there could be mistakes in the data set, perhaps introducing bias or duplication, or maliciously poisoned.   We don't discuss this further in this paper, but dependability of learning could benefit from its own treatise.  Moreover, it is possible that the run-time system relies on reinforcement learning, in which case the learning integrates with run-time, which will need to be considered separately.  

It is also possible that software bugs or other faults in the algorithm or supporting systems impact the error in accuracy.  That is, the affect of the software or hardware fault is visible in the accuracy of the AI part of the service.  Certain hardware errors may be prevalent only for GPUs that are used for AI.  All such cases would also be important to study, but we do not consider this type of `secondary' faults (from an AI perspective) in this paper.  

In summary, we suggest to add {\em AI Faults} under {\em Dimension} and introduce {\em AI Output Faults} as a fault class to further distinguish between {\em Non-ambiguous Faults} and {\em Semantic Ambiguity Faults}.  One could further distinguish fault classes, which will be useful for certain applications of AI.  Finally, we note that \cite{Avizienis04} is the result of several decades of discussions, and the proposal in this paper would similarly need to be considered and discussed by others.  Our main conclusion is that the fault classes do not naturally fit AI and could be modified to better feed into our understanding of dependability in AI-based services.   

\begin{figure}[htbp]
  \centering
  \includegraphics[width=0.45\textwidth]{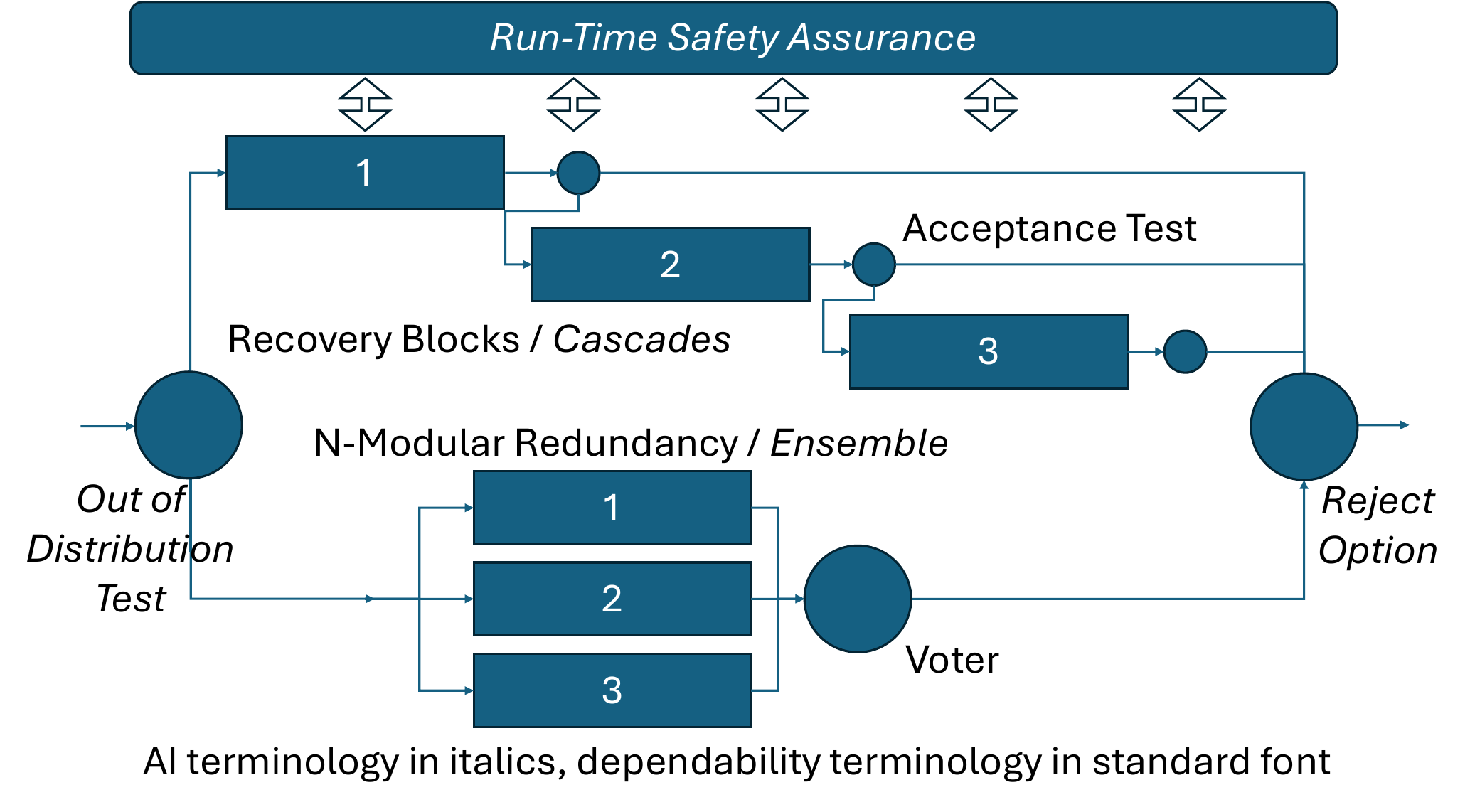}
  \caption{Fault Tolerance techniques from dependability and Accuracy Enhancement mechanisms from AI.}
  \label{fig:FT}
\end{figure}
\section{Fault Tolerance Techniques and their AI Counterparts}
\label{s:FT} 
We classify techniques according to Fig. \ref{fig:FT}, where the boxes are (copies of) AI algorithms implemented in software.  Data is fed into the algorithms, and follows the flow depicted through the arrows: left is the input, and right is the output.  The circles are decision operations, either a form of acceptance test, or a voter for ensemble techniques.  It terms of failures, we are interested in the output of algorithms, whether these are correct--if not, we consider this a failure. As we will see, there are techniques that reject the input or the output, and we do not consider rejection as a failure.  Instead, we assume that deselecting inputs or outputs is tolerated appropriately as part of the mechanism. 

We first reflect in Section \ref{s:cascades} and \ref{s:ensembles} on perhaps the two most basic and important fault tolerance techniques, namely Recovery Blocks and N-Modular Redundancy, and their counterparts in AI: Reject Option with Cascades and Ensembles. These techniques are essentially identical in traditional systems and AI, despite having significantly different motivation and justification when applied for dependability and AI accuracy, respectively.  We then discuss a number of safety/dependability techniques that are based on monitoring and execute controlling actions to maintain a level of safety or dependability.  These start from input and output rejection (Section \ref{ss:OOD} and \ref{ss:reject}) and culminate in full-blown Run-Time Safety Assurance systems (Section \ref{s:controller}).   

We note that our mapping at its core has been for techniques found for AI classifiers, and some of the ones we mention have slightly different names or designs in other uses of AI. 

\subsection{Ensembles (N-Modular Redundancy)}
\label{s:ensembles}
Introducing redundant components is an age-old approach to fault tolerance \cite{Avizienis1977NVersion}.  

{\bf N-Modular Redundancy.\ } In dependability, one of the most standard manners of providing fault tolerance is that of adding redundant copies, and then vote to establish the outcome.  If there are $N$ copies, this is called N-Modular Redundancy (NMR), and with 3 copies, it is called triple modular redundancy (TMR).  When the modules are software modules, the method is termed N-Version Programming (NVP).  In NMR, a voter decides about the outcome, typically based on majority voting. As long as a majority of the modules provides the correct outcome, the outcome of NMR is correct.  As reference for the work on software fault tolerance, we refer to the 1995 book {\em Software Fault Tolerance} edited by Michael Lyu \cite{Lyu95Book}, which contains chapters on all fault tolerance mechanisms mentioned here. 

One can argue that it is intuitively obvious that NMR is likely to improve the situation.  If it is rare that a component fails, then with three or more copies, it will be almost impossible for the majority to fail. In hardware fault tolerance, where it is highly unpredictable how faults will manifest themselves, this type of redundancy is highly effective. For software, it is harder to establish effective modular redundancy, since either the software or the inputs must be different from each other (the latter not always being possible). To produce multiple copies of software is usually prohibitively expensive, and even if different teams code different pieces of software, the faults may end up manifest themselves in the same `tough' parts of the code, failing on the same inputs.  In that case, redundant copies do not help, they all fail in the same way. 

{\bf AI Ensembles.\ }  In AI systems, the counterpart of NMR is captured as {\em ensembles} \cite{Dietterich2000Ensembles}.  Ensembles simply run in parallel a number of AI services for the same task, and then conduct a majority vote to decide the outcome.  For example, assume an AI classifier Each input image is classified by each of the classifiers, which deliver their outcome to a software-implemented voter, where the majority decides.  

Other than for NMR, it may not necessarily intuitively obvious whether an AI ensemble improves the situation: how can a number of inferior AI algorithms do better than relying on the one algorithm that is known to be the best?   The core argument in favour of ensembles was provided two and a half century ago by Condorcet's extremely impressive essay \cite{Condorcet1785Essai}. Condorcet provided us with the Jury Theorem as well as with the underpinning probabilistic reasoning. The Jury Theorem essentially explains why a voting jury, under certain conditions, can be shown to be preferred over a single judge, despite the poorer judgment of each individual juror. 

There are different ways to manipulate training to establish ensembles.  For instance, one can change the training data and create multiple classifiers, each with different training data. This approach works well for neural networks, because they tend to be very sensitive to the training data. Boosting and bagging \cite{Dietterich2000Ensembles} are variants of this approach.  One can also change the model, and it is also possible to manipulate the outputs to create an ensembles, e.g. by combining classes if the classifier has multiple classes \cite{Dietterich2000Ensembles}. Finally, one can inject randomness into the learning algorithm, and sample from the introduced randomness to create different versions of the learning.  In all these cases, the outcome is an ensemble of models, each trained in different manner.  

An excellent survey of the methods and basic approaches for AI ensembles can be found in Rokach' survey from 2010 \cite{rokach2010EnsembleSurvey}. Hansen \cite{Hansen1990NeuralNetworkEnsembles} in 1990 provided basic analytics to support and explain the working of ensembles, and Kucheva \cite{Kuncheva2003DiversityMeasures} provides grounding for diversity measures for ensembles.  The latter work is very interesting, the argument for the practical success of ensembles is not in independence, but in diversity of AI algorithms.  In fact, one can purposely introduce negative correlations between algorithms, e.g., through the used training data, to improve the outcome of the ensemble.  

The original motivation behind AI ensembles is quite different from that in fault-tolerance, but some of the same insights apply, particularly that the key to both is {\em diversity}.  In fault tolerance, the importance of diversity of the copies is easy to understand, since if the copies were identical, a software bug leading to a failure in one system will be leading to failure in the other versions as well.  In AI ensembles, the situation is more complex, and the work by Kucheva \cite{Kuncheva2003DiversityMeasures} provides a systematic study of diversity. For dependability, Littlewood and Miller \cite{Littlewood1989CoincidentFailures} provide a deep system analysis of N-Version Programming and its requirements for diversity. 

\subsection{Cascades (loosely related to Recovery Blocks)}
\label{s:cascades}

The basic working of the Recovery Block is that different service implementations are available, which are of different quality level \cite{Randell1975RecoveryBlocks,Lyu95Book}.  Starting from the highest quality implementation, a {em Acceptance Test} decides whether the result is sufficient.  If it is consider insufficient, the second implementation is executed, as depicted in the top half of Fig. \ref{fig:FT}.  This continues until all recovery blocks have been tried, in which case the system should go into a safe state (possibly implying control moves to a human operator). 

The effectiveness of the Recovery Block approach depends very much on the Acceptance Test.  We will see that this is related to the Reject Option, discussed in Section \ref{ss:reject}, where the decision to reject can be made by an AI module itself, for instance learned together with the main AI algorithm.  An Acceptance Test may be able to identify failure behaviour, for instance through corrupted output.  However, building a good Acceptance Test will become more complicated if failure behaviour cannot be recognised.  This is the case for the output of AI classifiers, since if the computer-implemented Acceptance Test would know if the outcome is correct, the AI classifier would not have been needed.  As an alternative, the AI algorithm may contain hyperparameters that provide information about the level of accuracy the output has.  For instance, the values of transformer token probabilities can be used as an indication of correctness of a Large Language Model \cite{KangBatu2025UncertaintyQuantificationHallucination} (leaving alone the ambiguity in determining what `correct' actually is, see Section \ref{s:dependability}).  

{\bf Cascades.\ } To create the counterpart of Recover Blocks in AI, one would combine the idea of Cascades \cite{ViolaJones2001Cascade} with the Reject Option technique of Section \ref{ss:reject} to introduce an explicit Acceptance Test.  The intend of a Cascade is for performance improvement, in which reliability must be maintained. Similar to Recovery Blocks, a series of AI algorithms is available to be executed, but different from Recovery Blocks, the outcomes of the algorithms `accumulate' until the outcome is acceptable.  In this manner, on aims to make quicker decisions if the outcome is certain enough, and one runs the algorithm longer if there is remaining uncertainty. This makes for very interesting performance reliability tradeoffs. We note that we are not aware of the use of Cascading for dependability alone.  

\subsection{Out of Distribution Detection}
\label{ss:OOD}
Out of Distribution (OOD) detection \cite{YangZhouLiLiu2024GeneralizedOODSurvey} aims to detect inputs that are, as the name says, out of distribution and therefore should not be fed into the classifier. In this manner, it avoids to run an algorithms for an input for which it may not be particularly suitable. Some of the formal verification developments of recent times can be leveraged in the decision about inputs, by considering whether inputs are within an envelope for which correctness can be guaranteed \cite{Liu19AlgorithmsVerifyingDNNs}. 

 To offer OOD detection capabilities is possibly complex in its own right, and most of the approaches are based on AI itself, to detect outliers or anomalies.   It is important to specify what happens if the input is withdrawn, otherwise one could consider omitting an output also as a failure.  The standard approach comes from the area of autonomous vehicles, in which one would want the human driver to take over if the driving systems detects an object it has never seen before \cite{YangZhouLiLiu2024GeneralizedOODSurvey}.  In this setting OOD detection governs when the system is allowed to act autonomously.  

\subsection{Reject Option}
\label{ss:reject}
Reject Option aims to reject the outcomes when the classifier is not confident the outcome was right \cite{Geifman2017SelectiveClassificationDNNs}.  In Fig. \ref{fig:FT} it is depicted on the righ-hand side, when the algorithms arrive at generating an output.  Chow introduced the idea of a reject option in 1957 \cite{Chow1957RejectOption}, while in their paper from 1969 \cite{Chow1970RejectOption} they provide a very clean analysis of this situation.  The idea behind the reject option is that when it is triggered, the human takes over, and in that sense it is the same as Out of Distribution detection, discussed in Section \ref{ss:OOD}.  However, it does not discard until after the classification takes place, and is therefore able to use knowledge in the decision making gained from the algorithm.  

An implementation of Reject Option in neural networks is discussed by Geifman and El-Yaniv \cite{Geifman2017SelectiveClassificationDNNs}.  In their proposal, the rejection function is learned together with the classifier, which leads to a guaranteed bound to the classification error while guaranteeing coverage, that is, the probability inputs would not be rejected.  One may consider different tradeoffs, which for instance would take into account the cost of rejecting an output and the cost of accepting an invalid output with the gain of accepting correct outputs \cite{deStefano2000SelectiveClassification}.   

The Out of Distribution detection and Reject Option techniques can also be combined with a AI-based monitor, that is, a Run-Time Safety Assurance system, see Section \ref{s:controller}.  

\subsection{AI-based Run-Time Safety Assurance}
\label{s:controller}
Can AI make AI reliable?  The answer is affirmative, and this set of techniques may be the most important, as well as the most challenging to asses (and therefore trust). We call this (set of) approaches Run-Time Safety Assurance, depicted at the top of the schema in Fig. \ref{fig:FT}.  

We have already seen in Reject Option that an AI rejection function can be learned together with a classifier or other AI algorithm, in such a manner that a combined risk function is optimised.  The idea behind run-time safety assurance is similar to a control system \cite{Lan2024ControlNN}, attempting to keep the overall systems within predefined bounds of performance.  The Run-Time Safety Assurance system gains sensory information from the AI service, monitors the AI service, and takes action if it finds a problem corresponding to out of bounds values of the sensors.  In this situation, it needs to decides what happens, and in many cases the system assumes a human to take control if the AI-based system exceeds predefined bounds.  

The Run-Time Safety Assurance system would typically rely on AI itself, and it would integrate a number of accuracy assurance techniques, such as Out-of-Distribution detection and Reject Option. Note that also in standard cyber physical control systems, the controller itself often actually is an AI-based system, e.g., a neural network \cite{Lan2024ControlNN}. The dependability design from Bloomfield and Rushby \cite{bloomfield2024dependability} adopts this idea as well, and a variety of architectures have been proposed in the literature.  An example is the integrated dependability cage \cite{Aniculaesei2023ConnectedDependabilityCage}, which combines a monitoring system with mechanisms that guarantee safety.  In general, the 'cage' idea provides a monitoring, AI-enabled, component that assures the performance of the AI system remains within predefined safety bounds. 

\section{Conclusion}
\label{s:conclusion}
In this paper we integrated AI-based systems and accuracy enhancing mechanisms for AI into the dependability taxonomy and traditional fault tolerance mechanisms.  There exist striking overlaps fault tolerance mechanisms and accuracy enhancing methods in AI.  N-Modular Redundancy finds it counterpart in ensembles, and the acceptance test in recovery blocks is similar to the reject option in AI.  This is perhaps somewhat surprising, since there are clear differences between aiming for dependability and increased accuracy.  We also integrate AI fault classes in the fault classes provided in the dependability taxonomy.  AI faults in outputs of algorithms are different in that the correctness of outputs is inherently uncertain. Integrating AI accuracy considerations into dependability will be critical in developing the design methods and techniques for dependable AI-based systems.  

\bibliographystyle{plain}
\bibliography{harms}

\end{document}